\documentclass[doublecol]{epl2} 
\usepackage{subfigure}

\begin{document}

\title{Nonequilibrium relaxation and scaling properties of the two-dimensional
       Coulomb glass in the aging regime}
\shorttitle{Nonequilibrium relaxation of the two-dimensional Coulomb glass}

\author{Matthew T. Shimer, Uwe C. T\"auber, \and Michel Pleimling}
\shortauthor{M. Shimer \etal}

\institute{Department of Physics, 
	   Virginia Polytechnic Institute and State University, 
	   Blacksburg, VA 24061-0435}


% to appear in Europhys. Lett. (2010); [arXiv:1007.1929].

\abstract{We employ Monte Carlo simulations to investigate the two-time density
autocorrelation function for the two-dimensional Coulomb glass.
We find that the nonequilibrium relaxation properties of this highly correlated
disordered system can be described by a full aging scaling ansatz.
The scaling exponents are non-universal, and depend on temperature and charge
density.}

\maketitle

\section{Introduction}

Exploring the nature of relaxation from out-of-equilibrium initial states
towards stationary thermal equilibrium has been a major focus of nonequilibrium
statistical mechanics (for a comprehensive current overview, see, e.g., 
Ref.~\cite{b.n}).
Fundamental questions concern the scaling properties of two-time correlation 
functions in the so-called aging regime (to be discussed more extensively
below), namely: 
(1) What are the basic scaling forms to describe numerical and experimental
data?
(2) Are the ensuing scaling exponents and scaling functions universal? 
(3) Which physical properties determine the scaling features in the aging 
regime?
Aside from contributing to our still rather incomplete understanding of 
processes far from equilibrium, providing comprehensive and reliable answers to
questions (2) and (3) could have important ramifications to materials science 
as well.
For example, a clear picture of which features of nonequilibrium relaxation 
processes contain specific information on the system under investigation in 
contrast to which properties are generic may provide a novel method of sample 
characterization.
To this end, various model systems have been carefully investigated through
numerical simulations; a few simple models even allow for analytical treatments
\cite{b.n}.

Disordered semiconductors in the vicinity of a metal--insulator transition
constitute an intriguing experimental system displaying complex relaxation
phenomena.
Relaxation measurements for the conductivity of a two-dimensional silicon 
sample have yielded unambiguous evidence for aging effects \cite{b.a}, see also
Ref.~\cite{epb}. 
The interpretation of these experiments continues to build upon the early 
theoretical predictions of Coulomb glass models \cite{b.o,b.u,b.l} that 
incorporate the essential physical aspects of charge carriers localized at 
random positions in space and interacting through long-range forces. 
Research into the non-trivial equilibrium properties 
\cite{b.c,b.d,b.e,b.v,b.f,c.d,c.e,c.c,pal} and non-equilibrium relaxation 
phenomena \cite{b.g,b.h,c.b,b.i,b.j,b.k,amir,c.a,ber} of the Coulomb glass over
the past two decades have considerably advanced our understanding of this 
paradigmatic model system for highly correlated disordered materials.
Our goal in this paper is to examine the fundamental questions (1)--(3) for the
Coulomb glass in two dimensions, with the aim to better understand universality
aspects and scaling properties of disordered semiconductors.
(We note that a very similar model, with the Coulomb potential essentially 
replaced by a logarithmic repulsion, describes the Bose glass phase of magnetic
flux lines in type-II superconductors that are pinned to columnar defects, see
Refs.~\cite{b.p,b.q,b.r}.)
Specifically, we address the detailed scaling form for the two-time density
autocorrelation function, and the dependence on temperature and carrier density
of the non-equilibrium relaxation kinetics.
Our work builds upon and extends the investigations of Grempel {\em et al.} who
employed a very similar Monte Carlo simulation method \cite{b.j,b.k}; however, 
we decided to extend our studies to multiple carrier densities in addition to 
varying temperatures. 
This allows us to study in a very systematic way the aging properties of the 
two-dimensional Coulomb glass, utilizing different scaling forms for the 
density autocorrelation function (see also Ref.~\cite{c.f} in the context of a 
spin glass model).

\section{The Coulomb glass model and Monte Carlo simulation procedure}

Our basic model system is a Coulomb glass in two dimensions.
The Coulomb glass model, introduced by Efros and Shklovskii \cite{b.o}, 
consists of multiple localized pinning sites available to the charge carriers. 
Because of the strong intra-site correlations these sites can only contain a 
single charge carrier at most. 
The system is dominated by long-range repulsive Coulomb interactions 
$V(r) \sim 1/r$ and the spatial disorder is induced by the randomly located (on
a continuum) available sites. 
The Hamiltonian of the Coulomb glass model reads \cite{b.o,b.u,b.l}
\begin{equation}
\label{eq.1}
	\displaystyle H = \sum_i n_i \varphi_i + \frac{e^2}{2 \kappa} 
	\sum_{i\neq j} \frac{(n_i-K)(n_j-K)}{| {\bf R}_i - {\bf R}_j |} \, ,
\end{equation}
where ${\bf R}_i$, $\varphi_i$, and $n_i$ respectively denote the position 
vector (here in two dimensions), (bare) site energy, and occupancy of the $i$th
site, $i = 1,\ldots,N$.
The occupancy $n_i$ can only take on the values $0$ or $1$.  
The first term corresponds to (random) site energies assigned to each 
accessible location; since the system is dominated by the long-range forces, we
chose $\varphi_i = 0$ to further simplify the model, while still allowing the 
positions ${\bf R}_i$ to be continuous and random \cite{b.e,b.h,b.j}. 
(Alternatively, one could allow $\varphi_i$ to take on random values while the
positions ${\bf R}_i$ would be discrete and set on a lattice.)
The second contribution encapsulates the repulsive Coulomb interactions (with
dielectric constant $\kappa$).
In order to maintain global charge neutrality, a uniform relative charge 
density $K = \sum_i n_i / N$ is inserted; $K$ can also be described as the 
total carrier density per site, or filling fraction.
We note that with $\varphi_i = 0$ the Hamiltonian (\ref{eq.1}) displays
particle--hole symmetry, i.e., systems with $K = 0.5 + k$ and $K = 0.5 - k$ are
equivalent.
Upon replacing the occupation numbers with spin variables 
$\sigma_i = 2 n_i - 1 = \pm 1$, one sees that the Coulomb glass maps onto a 
random-site random-field antiferromagnetic Ising model with long-range exchange
interactions \cite{b.c}.

We employ a Monte Carlo simulation algorithm to at least heuristically model 
the kinetics in the Coulomb glass \cite{b.j,b.k}.
At each time step, one randomly selected charge carrier attempts to hop from an
occupied site $a$ to an empty site $b$. 
Two factors determine the hopping success rate, namely a strongly 
distance-dependent tunneling process (in semiconductors mediated through
phonons) and thermally activated jumps over energy barriers \cite{b.j}, i.e.,
\begin{equation}
\label{eq.2}
	\Gamma_{a \rightarrow b} = \tau_0^{-1} e^{- 2 R_{ab} / \xi} \,
	\tx{min}[1,e^{- \Delta E_{ab} / T}] \, ,
\end{equation}
where $\tau_0$ represents a microscopic time scale, 
$R_{ab} = |{\bf R}_a - {\bf R}_b|$ is the distance between sites $a$ and $b$,
$\xi$ characterizes the spatial extension of the localized carrier wave 
functions, and we have set $k_B = 1$.
The first, spatially exponential term in (\ref{eq.2}) is derived from quantum 
tunneling, while the second exponential term is due to thermodynamics:
A thermally activated hop from sites $a$ to $b$ entails the energy difference
\begin{equation}
\label{eq.3}
	\Delta E_{ab} = \epsilon_b - \epsilon_a - \frac{e^2}{\kappa R_{ab}}\, ,
\end{equation}
with the (interacting) site energies
\begin{equation}
\label{eq.4}
	\epsilon_a = \frac{e^2}{\kappa} \displaystyle \sum_{b\neq a}
      	\frac{n_b-K}{R_{ab}} \, .
\end{equation}
In the following, distances are measured relative to the average separation 
between sites $a_0$, and energies as well as temperature scales are measured 
relative to the typical Coulomb energy $E_C = e^2 / \kappa a_0$ \cite{b.j}.  
We set the spatial extension of the localized wave functions, $\xi$, to $a_0$,
as in Refs.~\cite{b.j,b.k}. 
We have explored other values for $\xi$ as well, and (within the applicability
range of the model) found the ensuing results to simply scale with $\tau_0$; 
hence $\xi = a_0$ was chosen for simplification.

The Monte Carlo simulations were initiated by randomly placing $N$ sites within
a square simulation cell of length $L$ containing $N_e = K N$ charge carriers,
where $N = L^2$. 
We performed simulations for systems with $L=8, 10, 16, 32$; with temperatures 
in the range of $0.01 \le T \le 0.05$; and carrier densities in the range of 
$0.375 \le K \le 0.5$ (also equivalent to $0.5 \leq K \leq 0.625$ due to 
particle--hole symmetry). 
Running the simulations with various system sizes $L$, we noticed no measurable
finite-size effects, as demonstrated in Fig.~\ref{fig.2a}.
Temperatures larger than $0.03$ turned out not to be useful for our study of
aging processes since equilibrium was then reached far too quickly.
In contrast, for $T < 0.01$, the kinetics slowed down too much for gathering 
statistically significant data within computationally reasonable time frames.
As will be discussed in more detail below, the dynamics also freezes out within
the numerically accessible simulation times for $K < 0.4$ (or $K > 0.6$).

Periodic boundary conditions were used and the potential due to charges outside
the cell was calculated by mirroring the simulation cell on each side.  
Initially, the charge carriers were placed randomly at available sites to 
mimic a quench from very high temperatures. 
Then the system was evolved at temperature $T$ with the dynamics described by 
the generalized Metropolis rate (\ref{eq.2}).  
The thermally activated, tunneling-controlled (variable-range) hopping 
algorithm begins with randomly choosing an occupied site $a$ and assigning a 
normalized probability proportional to $\exp( - 2 R_{ab} / \xi)$ to each empty 
site.  
An empty site $b$ is then chosen from this probability distribution and a hop 
from $a$ to $b$ is attempted.  
The success probability for this move is determined by 
$\tx{min}[1,e^{- \Delta E_{ab} / T}]$.  
There are $N$ hop attempts for each Monte Carlo step (MCS).  
Our simulation runs consisted of $1 \times 10^6$ MCS and results were averaged 
over $3000$ different realizations of the initial conditions and of the 
random number sequences.

\begin{figure}
\onefigure[width=8.5cm]{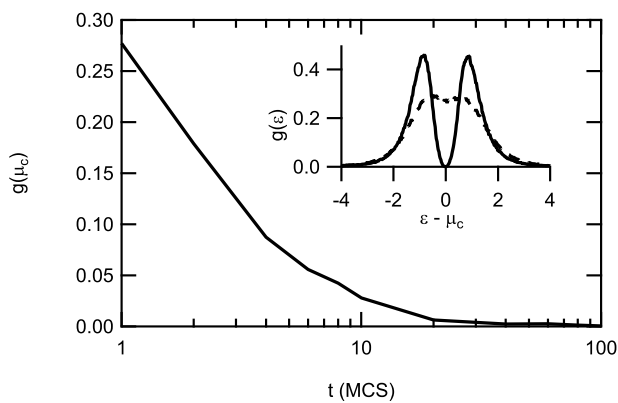}
\caption{Coulomb gap formation in a Coulomb glass ($L = 16$, $N = 256$ sites) 
	at half filling $K = 0.5$ and at temperature $T = 0.02$.
 	The plot shows how $g(\mu_c)$ quickly approaches zero.
	The density of states is displayed in the inset at $t = 1$ MCS 
	(dashed) and $t = 100$ MCS (solid), when the Coulomb gap correlations 
	are clearly established.}
\label{fig.1}
\end{figure}

In thermal equilibrium, the long-range correlations and spatial disorder 
conspire to produce a soft gap in the (interacting) density of states 
$g(\epsilon)$, as famously first noted by Efros and Shklovskii \cite{b.o}.
The repulsive interactions induce strong spatial anticorrelations, which in 
turn suppress the availability of states with energies close to the chemical 
potential $\mu_c$ that at $T = 0$ separates the occupied and empty energy 
levels.
The ensuing Coulomb gap appears to asymptotically be described by a power law
$g(\epsilon) \sim |\epsilon - \mu_c|^s$ 
\cite{b.u,b.l,b.c,b.d,b.e,b.v,b.f,c.d,c.e,c.c}; within a simple mean-field 
approach, $s = (d / \sigma) - 1$ in $d$ dimensions for a system with long-range
repulsive potential $V(r) \sim 1 / r^\sigma$.
We have monitored the emergence of the Coulomb gap in our Monte Carlo 
simulations.
Quite remarkably, the soft gap forms very fast, as illustrated in 
Fig.~\ref{fig.1}, which shows how the density of states at the chemical 
potential $g(\mu_c)$ quickly approaches zero.
Indeed, the Coulomb gap is already pronounced after as few as $10$ MCS, and 
clearly established in less than $100$ MCS.
In the following, we proceed to carefully analyze the slow decay of local
density correlations; we remark that the interesting aging kinetics happens in
a time period when the Coulomb gap itself is already well-established and
therefore addresses slow dynamics in a highly correlated disordered system.

\section{Scaling properties or the two-time density autocorrelation function in
the aging regime}

In the following, we focus on the (normalized) two-time carrier density 
autocorrelation function \cite{b.j}, 
\begin{equation}
\label{eq.5}
	C(t,s) = \frac{\langle n_i(t) n_i(s) \rangle - K^2}{K (1-K)} = 
	\frac{\sum_i n_i(t) n_i(s) - N K^2}{N K (1-K)} \, ,
\end{equation}
where $t > s$, and the waiting time $s$ refers to the time elapsed since the 
high-temperature quench (initiation of the Monte Carlo simulations). 
If both $s$ and $t$ are large compared to microscopic time scales $\tau_0$, and
also well separated, i.e., if
\begin{equation}
\label{eq.6b}
	t \gg \tau_0 \, , \quad s \gg \tau_0 \, , \quad t-s \gg \tau_0 \, ,
\end{equation}
time-translational invariance is broken and $C$ depends on both $t$ and $s$ 
separately.
In the aging scaling regime, one then expects the autocorrelation function to
obey the general scaling form (we follow the notation in Ref.~\cite{b.n}),
\begin{equation}
\label{eq.6}
	C(t,s) = s^{-b} \, f_C(t/s^\mu) \, .
\end{equation}
For certain simple systems obeying purely relaxational dynamics, namely the 
one-dimensional kinetic Ising model \cite{God,Lip}, the time-dependent 
Ginzburg--Landau models quenched to a critical point, \cite{b.w,b.x} and the 
spherical model A coarsening dynamics in the low-temperature phase, with 
short-range \cite{b.y,b.z} or long-range \cite{Can,Bau,Dut} interactions, the 
aging scaling regime is amenable to analytic treatment.
In these cases one finds that the ansatz (\ref{eq.6}) is satisfied with 
exponent $\mu = 1$, a situation commonly referred to as full aging scaling.
For more complex systems such as spin glasses the possibility of a subaging 
scaling with $\mu < 1$ \cite{b.n} has been discussed in the literature.
For $\mu = 1$ and $t/s \rightarrow \infty$ the full aging scaling function 
follows a power law \cite{b.n},
\begin{equation}
\label{eq.7}
	f_C(t/s) \sim (t/s)^{-\lambda_C/z} \, ,
\end{equation}
with the autocorrelation exponent $\lambda_C$ and the dynamic exponent $z$.
Overall, this presents us with three scaling exponents: $b$, $\mu$, and 
$\lambda_C/z$.

From the averaged charge density autocorrelation function (\ref{eq.5}) we 
extracted the scaling exponents using an interpolation method motivated by 
Ref.~\cite{b.m}. 
The total variance, or spread, of the data was calculated by comparing the data
points to polynomial interpolations of all other curves. 
This variance was only measured within the scaling regime. 
We have chosen to either assume full aging scaling, $\mu = 1$ in 
Eq.~(\ref{eq.6}), and applied the polynomial interpolation method to obtain the
scaling exponent $b > 0$; or to set $b = 0$ (simple scaling ansatz) and
alternatively determine the subaging exponent $\mu < 1$.

\begin{figure}
\subfigure[]{\label{fig.2a}\includegraphics[width=8.0cm]{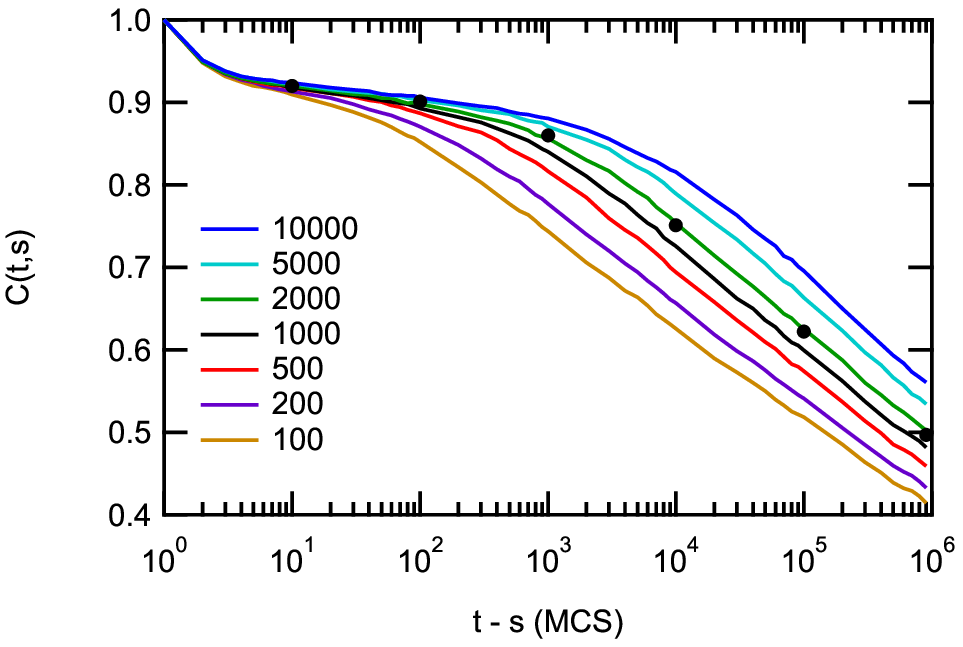}}
\subfigure[]{\label{fig.2b}\includegraphics[width=8.0cm]{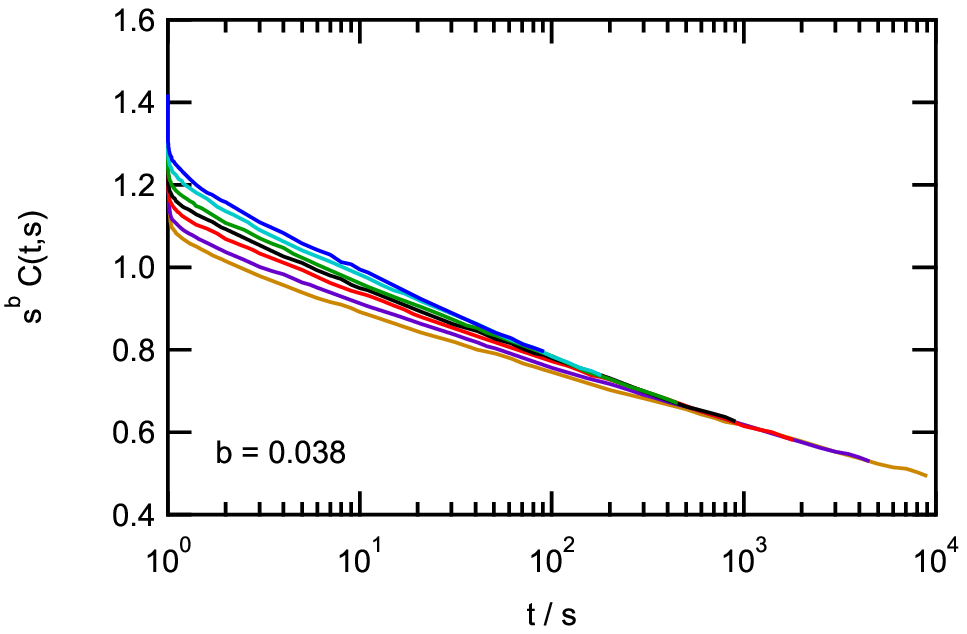}}
\subfigure[]{\label{fig.2c}\includegraphics[width=8.0cm]{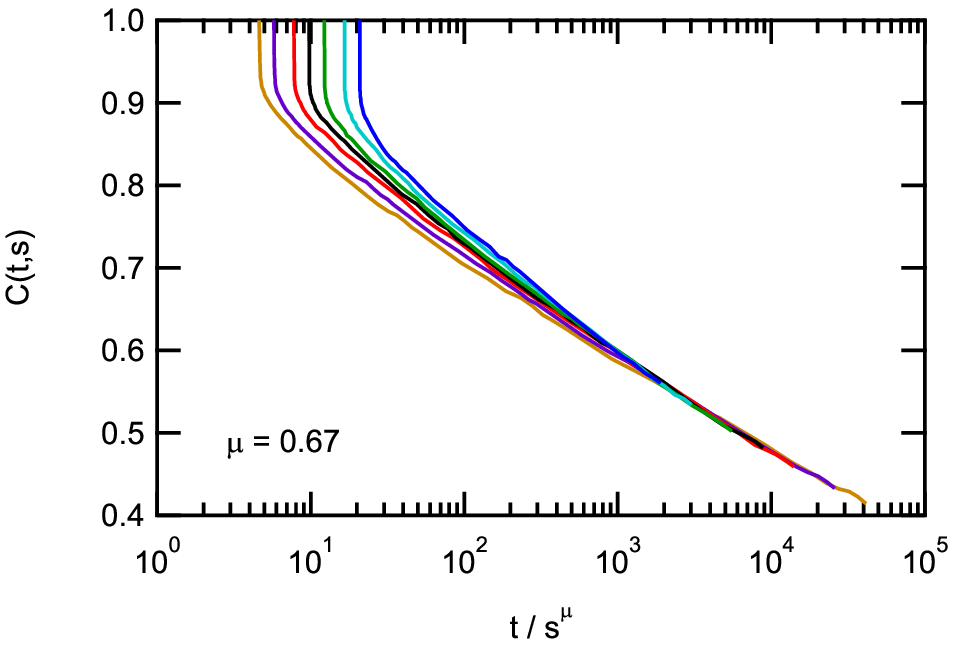}}
\caption{Carrier density autocorrelation function $C(t,s)$ at half filling 
	$K = 0.5$, in a system of linear extension $L = 16$ ($N = 256$ sites) 
	at temperature $T = 0.02$, measured for various waiting times 
	$s=10^2,2 \times 10^2, 5 \times 10^2, 10^3, 2 \times 10^3, 
	5 \times 10^3, 10^4$ (from bottom to top).
	(a) The plot vs. $t-s$ demonstrates the breaking of time-translation
	invariance. The $\bullet$ symbols represent data obtained from a system
        of length $L = 10$ ($N = 100$ sites).
	(b) A full aging scaling plot, which assumes $\mu = 1$, yields the 
	scaling exponents $b = 0.038$ and $\lambda_C/z = 0.103$. 
	(c) A subaging scaling analysis, fixing $b = 0$, gives $\mu = 0.67$.}
\label{fig.2}
\end{figure}
\begin{figure}
\subfigure[]{\label{fig.3a} \includegraphics[width=8.0cm]{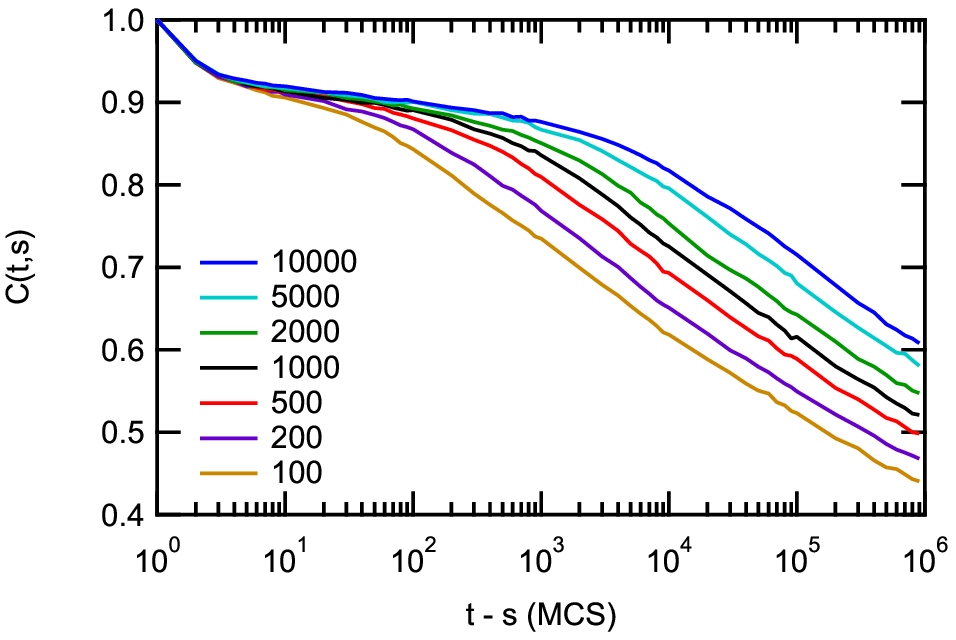}}
\subfigure[]{\label{fig.3b} \includegraphics[width=8.0cm]{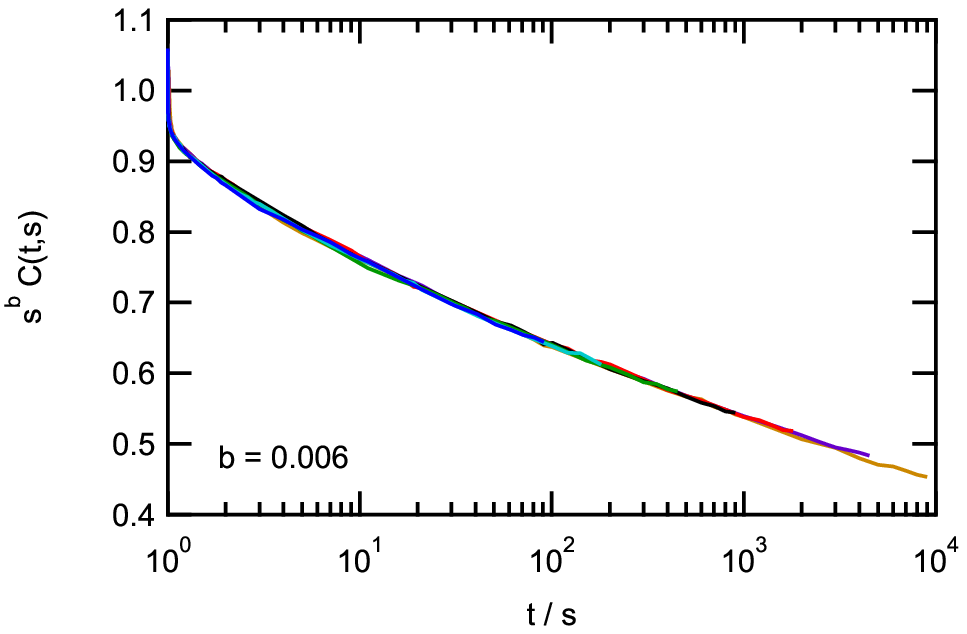}}
\subfigure[]{\label{fig.3c} \includegraphics[width=8.0cm]{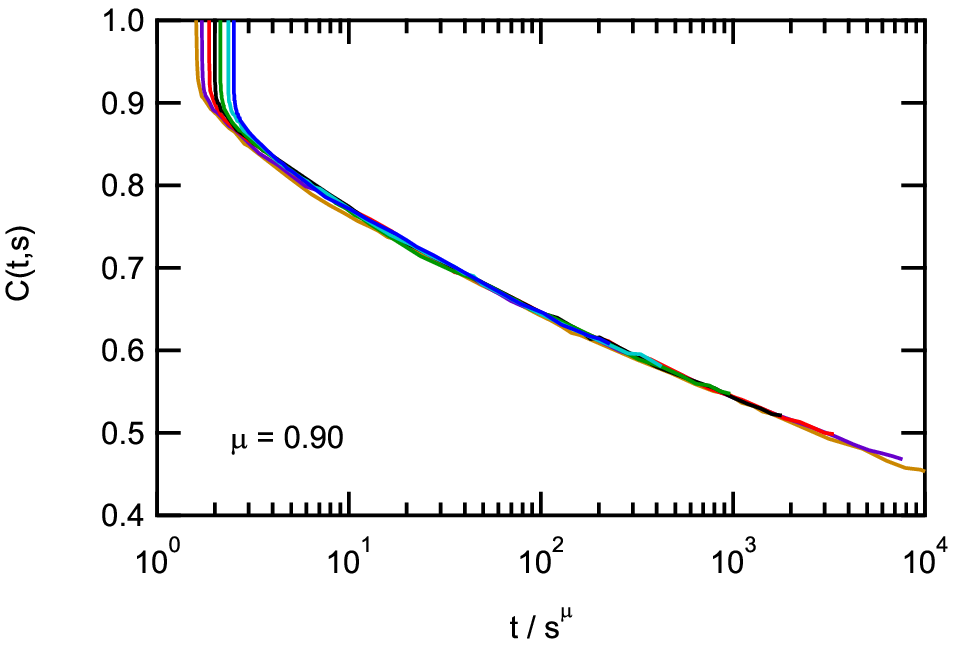}}
\caption{(a) Carrier density autocorrelation function $C(t,s)$ at carrier 
	density $K = 0.4375$ (again with $L = 16$, $N = 256$, and $T = 0.02$), 
	measured for various waiting times $s = 10^2, 2 \times 10^2, 
	5 \times 10^2, 10^3, 2 \times 10^3, 5 \times 10^3, 10^4$ (from bottom 
	to top). Note the markedly slower decay as compared to the date at half
	filling shown in Fig.~\ref{fig.2}. 
	(b) A full aging scaling plot ($\mu = 1$) gives $b = 0.006$ and 
	$\lambda_C / z = 0.075$.  
	(c) A subaging scaling analysis ($b = 0$) results in $\mu = 0.90$.}
\label{fig.3}
\end{figure}
Sample sets of results (with $L = 16$, $N = 256$ sites) are shown in 
Figs.~\ref{fig.2} and \ref{fig.3}. 
The density autocorrelation data at half filling $K = 0.5$ displayed in 
Fig.~\ref{fig.2a} match nicely with the results of Ref.~\cite{b.j}; however, 
the scaling forms used in Figs.~\ref{fig.2b} and \ref{fig.2c} are different 
from those employed by Grempel. 
As is apparent from Fig.~\ref{fig.2a}, the density autocorrelation function is
characterized by two temporal regimes. 
In the first time regime, the density autocorrelation relaxes towards a plateau
(often referred to as $\beta$ relaxation in the glass literature), whereas in
the second time regime the system very slowly approaches equilibrium ($\alpha$
relaxation) \cite{gl1,gl2}.
This later regime displays breaking of time translation invariance and aging
scaling, provided the inequalities (\ref{eq.6b}) hold. 
With our current algorithm and available hardware, we cannot effectively 
investigate systems with densities lower than $K = 0.40625$ and temperatures 
lower than $T = 0.02$, since in these configurations the plateau regions 
persist for much longer times than are computationally accessible for us, i.e.,
the system essentially freezes into a structure that corresponds to the $\beta$
relaxation plateau.
Notice that the autocorrelation data obtained for a smaller system ($L = 10$, 
$N = 100$ sites) coincide with those for the larger system, demonstrating the 
absence of finite-size effects in the observed time window.

In Fig.~\ref{fig.2b}, we plot the same data as in Fig.~\ref{fig.2a}, after 
applying a rescaling procedure that assumes full aging (where $\mu = 1$).
One notes that the systematic deviations for small $t$ become less and less
important for increasing waiting times $s$.
From the scaling ansatz (\ref{eq.6}), which is well founded theoretically 
\cite{b.n}, the scaling exponent $b$ can be reliably extracted from the data 
sets corresponding to larger waiting times.
As expected from Eq.~(\ref{eq.7}), a power law is observed at long times, which
in turn yields the scaling exponent $\lambda_C/z$.
At half filling $K = 0.5$, we find $b = 0.038 \pm 0.006$ and 
$\lambda_C / z = 0.103 \pm 0.004$.
Alternatively, we may employ a subaging scaling ansatz, setting $b = 0$.
Figure~\ref{fig.2c} shows the results from scaling using $\mu$ only.
Also in this case deviations are observed for $t$ small, but an improvement is 
less apparent when increasing the waiting time.
Collapsing the data on a single master curve at large scaling arguments works
apparently about equally well with either method employed in (b) or (c).
In fact, allowing nontrivial values for both scaling exponents, $b > 0$ and 
$\mu < 1$, yields comparable scaling collapse quality.
However, given that theoretical analysis (admittedly for comparatively simple 
systems) yields the full aging scenario with $\mu = 1$ and generally $b > 0$,
we would argue that subaging scaling, as used in previous studies 
\cite{b.j,b.k}, need not be invoked to describe the nonequilibrium relaxation 
phenomena in the Coulomb glass.

\begin{table}
\caption{Charge carrier density dependence of the full aging scaling exponents 
	$b$ and $\lambda_C / z$ (with $\mu = 1$ and $T=0.02$).}
\label{tab.1}
\begin{center}
\begin{tabular}{p{1.3cm} c c}
$K$ & $b$ & $\lambda_C / z$ \\ \hline \noalign{\smallskip}
0.40625 & \parbox[t]{2.5cm}{\raggedleft $0.001 \pm 0.002$} & $0.046 \pm 0.004$ 
\\
0.4375  & \parbox[t]{2.5cm}{\raggedleft $0.006 \pm 0.002$} & $0.075 \pm 0.004$ 
\\
0.46875 & \parbox[t]{2.5cm}{\raggedleft $0.021 \pm 0.004$} & $0.089 \pm 0.003$ 
\\
0.5     & \parbox[t]{2.5cm}{\raggedleft $0.038 \pm 0.006$} & $0.103 \pm 0.004$
\\ 
\end{tabular} 
%\end{center}
%\end{table}
%
%\begin{table}
\caption{Temperature dependence of the full aging scaling exponents $b$ and 
	$\lambda_C / z$ (with $\mu = 1$ and $K=0.5$).}
\label{tab.2}
%\begin{center}
\begin{tabular}{p{2cm} c c}
$T$ & $b$ & $\lambda_C / z$ \\ \hline \noalign{\smallskip}
0.01 & \parbox[t]{2.5cm}{\raggedleft $-0.001 \pm 0.001$} & $0.036 \pm 0.004$ \\
0.02 & \parbox[t]{2.5cm}{\raggedleft $0.038 \pm 0.006$}  & $0.103 \pm 0.004$ \\
0.03 & \parbox[t]{2.5cm}{\raggedleft $0.080 \pm 0.008$}  & $0.173 \pm 0.005$ 
\end{tabular} 
\end{center}
\end{table} 
\begin{table}
\caption{Charge carrier density dependence of the subaging exponent $\mu$ (with
	$b = 0$ and $T=0.02$).}
\label{tab.3}
\begin{center}
\begin{tabular}{p{2cm} c}
$K$ & $\mu$ \\ \hline \noalign{\smallskip}
0.40625 & \parbox[t]{2.5cm}{\raggedleft $1.00 \pm 0.01$} \\
0.4375  & \parbox[t]{2.5cm}{\raggedleft $0.90 \pm 0.01$} \\
0.46875 & \parbox[t]{2.5cm}{\raggedleft $0.78 \pm 0.01$} \\
0.5     & \parbox[t]{2.5cm}{\raggedleft $0.67 \pm 0.02$} \\ 
\end{tabular} 
%\end{center}
%\end{table}
%
%\begin{table}
\caption{Temperature dependence of the subaging exponent $\mu$ (with $b = 0$ 
         and $K=0.5$).}
\label{tab.4}
%\begin{center}
\begin{tabular}{p{2cm} c}
$T$ & $\mu$ \\ \hline \noalign{\smallskip}
0.01 & \parbox[t]{2.5cm}{\raggedleft $0.98 \pm 0.06$} \\
0.02 & \parbox[t]{2.5cm}{\raggedleft $0.67 \pm 0.02$} \\
0.03 & \parbox[t]{2.5cm}{\raggedleft $0.54 \pm 0.02$} 
\end{tabular} 
\end{center}
\end{table} 
Figure~\ref{fig.3} shows the results when the same scaling methods are applied 
to our data for relative charge carrier density $K = 0.4375$ (or $K = 0.5625$).
As the carrier density deviates from half filling, the relaxation becomes 
markedly slower.
%Notice that the data line in Fig.~\ref{fig.3b} that appears to not be properly
%scaled is the one for the shortest waiting time $s = 100$. 
%We interpret this as an example where $s$ is too small for the second 
%inequality in (\ref{eq.6b}) to be satisfied; hence the aging scaling ansatz
%does not apply. 
%Discarding this data set, 
The full aging scaling ansatz ($\mu = 1$) yields an excellent data collapse for
all $s$ and $t$ values. 
The same conclusion holds for other charge carrier densities $K \neq 0.5$.
For $K = 0.4375$, from the scaling analysis based on Fig.~\ref{fig.3b} we infer
the rather low value $b = 0.006 \pm 0.002$ along with 
$\lambda_C / z = 0.075 \pm 0.004$.
If we instead assume $b = 0$, the subaging scaling from Fig.~\ref{fig.3c} gives
$\mu = 0.90 \pm 0.01$.
Clearly, the aging scaling exponents are {\em not} universal in the Coulomb 
glass, but strongly depend on the charge carrier density.
At least near half-filling, in the time window that is accessible to us we
observe power-law scaling, albeit with numerically small exponent values $b$ 
and $\lambda_C / z$, different from, but close to the logarithmic dependence 
obtained by the mean-field type analysis in Refs.~\cite{c.a, amir}.

We ran sets of 3000 simulation runs each at various values of $T$ and $K$ to 
systematically explore the dependence of the aging scaling exponents on the
temperature (already noted in Refs.~\cite{b.j,b.k}) and carrier density.
The resulting exponent values for $b$ and $\lambda_C / z$ as obtained within 
the full aging framework (setting $\mu = 1$) are listed in Tables~\ref{tab.1} 
and \ref{tab.2}.
Physically, smaller values of the exponents imply slower relaxation kinetics. 
The trend that is observed is that as density drops away from half filling, the
relaxation slows. 
A similar trend is seen in the temperature dependence; namely, the relaxation
processes become slower at lower temperatures, as one would naturally expect.
If, on the other hand, we enforce $b = 0$ and instead allow for the subaging
scaling scenario, we find the exponent values $\mu$ listed in 
Tables~\ref{tab.3} and \ref{tab.4}. 
Note that a larger value of $\mu$ leads to slower relaxation.
The dependence of the subaging exponent thus follows the same qualitative 
trends as observed in full aging scaling.
Naturally, as in the full aging scaling analysis the exponent $b$ approaches
zero at low temperatures or away from half-filling, in the subaging scaling
analysis $\mu \to 1$.
(In fact, for either $T = 0.01$ or $K = 0.40625$ our simulations approach the
aforementioned limitations in run times; we list the corresponding exponent 
values with large relative errors nevertheless, since they still confirm the 
general trends.)
We remark that our findings for the scaling of nonequilibrium relaxation 
processes in the Coulomb glass align well with a recent study of the aging 
kinetics in the two-dimensional ferromagnetic Ising model with uniform bond 
disorder: Ref.~\cite{b.s} also finds non-universal scaling exponents, depending
on the ratio of disorder distribution width and temperature.

\section{Conclusion}

We have investigated nonequilibrium relaxation processes of the two-dimensional
Coulomb glass model at low temperatures via Monte Carlo simulations, and 
confirmed that its scaling features are governed by the simple aging scaling
form. 
Through the use of polynomial interpolation methods, three scaling exponents 
were extracted and found to follow a common trend: as either the temperature
decreases or the charge carrier density deviates more from half-filling, the 
exponents reflect slower relaxation kinetics. 
The aging scaling exponents are thus found to be non-universal, as is the case
for disordered magnetic systems \cite{b.s}.
Our results indicate that the inclusion of a subaging exponent $\mu \not= 1$ is
unnecessary since a nonzero scaling exponent $b$ encapsulates the same 
characteristics, and the corresponding simple aging scaling form is more firmly
grounded on theoretical analysis.

We are presently working on extending the scope of our study into three 
dimensions. 
Preliminary data for the density autocorrelation function scaling in the aging
regime suggest the same basic features and trends that we observe in two 
dimensions \cite{b.t}.
We also plan to implement logarithmic repulsion instead of the $1/r$ potential
used in the Coulomb glass, with the aim to study aging phenomena in the Bose 
glass phase of type-II superconductors with columnar defects 
\cite{b.p,b.q,b.r}. 
Our ultimate goal will be to more closely relate our observables and findings 
to experimental setups and measurements. 

\acknowledgments
This work was supported by the U.S. Department of Energy, Office of
Basic Energy Sciences (DOE--BES) under grant no. DE-FG02-09ER46613.
We gladly acknowledge stimulating discussions with Ariel Amir, Malte Henkel, 
and Dragana Popovi\'c.

\end{document}